\documentclass[10pt,onecolumn,aps,prd,preprintnumbers,showpacs,superscriptaddress,nofootinbib,amsmath,amssymb,floats,floatfix,showkeys,notitlepage,longbibliography]{revtex4-1}
\usepackage{orcidlink}
\UseRawInputEncoding
\usepackage{comment}
\usepackage{lipsum}
\usepackage{graphicx}
\usepackage{subfigure}
\UseRawInputEncoding
\usepackage{palatino}
\usepackage{sans}
\usepackage{hyperref}
\hypersetup{colorlinks=true,linkcolor=blue,urlcolor=blue,citecolor=blue}
\usepackage[toc,page]{appendix}
\usepackage[normalem]{ulem}
\usepackage{adjustbox}
\usepackage{latexsym}
\usepackage{amsmath}
\numberwithin{equation}{section}
\usepackage{amssymb}
\usepackage{amsfonts}
\usepackage{dcolumn}
\usepackage{bm}
\usepackage{tikz}
\usetikzlibrary{decorations.pathmorphing}
\usepackage{bigints}
\usepackage{array,tabularx,multirow,booktabs}
\usepackage[tracking=true]{microtype}
\usepackage{soul} %for highlighting
\SetTracking{}{500}
\SetTracking{encoding={*}, shape=sc}{40}
\UseRawInputEncoding %for inputenc error%
\allowdisplaybreaks

\usepackage[utf8]{inputenc}
\usepackage{xcolor} % For colored text if needed

\newcommand{\nn}{\nonumber}

\begin{document} \sloppy
\title{Noether charges and the first law of thermodynamics for multifractional \\ Schwarzschild black hole in the q-derivative theory}

\author{Reggie C. Pantig \orcidlink{0000-0002-3101-8591}} 
\email{rcpantig@mapua.edu.ph}
\affiliation{Physics Department, School of Foundational Studies and Education, Map\'ua University, 658 Muralla St., Intramuros, Manila 1002, Philippines.}

%\date{\today}

\begin{abstract}
In this paper, we investigate black-hole thermodynamics in the multi-fractional theory with $q$-derivatives, focusing on static, spherically symmetric vacuum solutions in the spherical-coordinate approximation. In the geometric frame the solution is exactly Schwarzschild in the areal radius $q$, so that canonical charges can be defined using standard covariant methods. The conserved mass depends only on the Schwarzschild integration constant, and the Iyer--Wald entropy satisfies the usual area law in terms of the geometric horizon radius. When the Hawking temperature is defined in the fractional radial coordinate $r$, however, it acquires an explicit dependence on the multi-fractional profile through the local factor $q'(r_{\rm h})$ at the horizon. As a result, variations of the non-dynamical profile parameters generically obstruct integrability of a naive Clausius relation expressed solely in terms of mass and entropy. We show that this obstruction is resolved by enlarging the thermodynamic state space to include the profile parameters and by constructing an integrable entropy functional obtained from a radial integral of the geometric radius. The corresponding extended first law contains additional work terms conjugate to the multi-fractional couplings. We analyze both binomial and log-oscillating profiles, clarify the role of presentation dependence, and delineate the consistency conditions required for a well-defined exterior branch with a single horizon. Our results make explicit the separation between profile-insensitive canonical charges and profile-sensitive thermal quantities in multi-fractional black-hole thermodynamics.
\end{abstract}



\maketitle

\section{Introduction} \label{sec1}
Black holes provide a uniquely sharp arena where the interplay between gravity, quantum physics, and spacetime structure becomes unavoidable. On the one hand, their classical mechanics is governed by a handful of geometric invariants; on the other hand, their thermodynamic behavior ties together horizon geometry, conserved charges at infinity, and quantum field theory in curved spacetime. The entropy--area connection was introduced by Bekenstein \cite{Bekenstein:1973ur}. The four laws of black-hole mechanics were formulated by Bardeen, Carter, and Hawking \cite{Bardeen:1973gs}. Hawking's particle-creation calculation established the thermal character of black-hole radiation \cite{Hawking:1975vcx}. The Euclidean action approach of Gibbons and Hawking then provided a gravitational path-integral framework for black-hole thermodynamics \cite{Gibbons:1976ue}. Because these elements probe both the infrared (astrophysical) regime and ultraviolet sensitivities (horizon-scale physics and beyond), black holes are natural laboratories for testing effective models of quantum-gravity-induced departures from general relativity.

A recurring theme across many quantum-gravity approaches is \emph{dimensional flow}, namely the emergence of an effective spacetime dimension that depends on the probed scale. Carlip gives a broad review of dimensional reduction in quantum-gravity models \cite{Carlip:2017eud}. Earlier, 't Hooft discussed dimensional reduction as a quantum-gravity principle \cite{tHooft:1993dmi}. In causal dynamical triangulations, scale-dependent spectral dimension was found by Ambjorn, Jurkiewicz, and Loll \cite{Ambjorn:2005db}, and was later used by Sotiriou, Visser, and Weinfurtner as a probe of the ultraviolet continuum regime \cite{Sotiriou:2011mu}. In asymptotically safe gravity, Lauscher and Reuter found fractal spacetime structure through renormalization-group methods \cite{Lauscher:2005qz}. Ho\v{r}ava's Lifshitz-point model provides another route to anisotropic ultraviolet scaling \cite{Horava:2009uw}.

Multi-fractional spacetimes were designed as a minimal, field-theoretic realization of dimensional flow, in which multi-scale properties are encoded directly in the integration measure and in the differential structure of the theory. A comprehensive review is given in Ref. \cite{Calcagni:2016azd}, with an updated account in Ref. \cite{Calcagni:2021ipd}. The early fractal-universe formulation introduced the basic multi-scale measure idea \cite{Calcagni:2009kc}. Its field-theoretic, gravitational, and cosmological development was given in Ref. \cite{Calcagni:2010bj}. The geometric formulation of multi-fractional spacetime was developed in Ref. \cite{Calcagni:2011sz}, while the associated geometry of fractional spaces was analyzed in Ref. \cite{Calcagni:2011kn}. Among the various realizations, the theory with $q$-derivatives (often denoted $T_q$) is technically simple and physically transparent: the dynamics is obtained by replacing the ordinary coordinates $x^\mu$ with \emph{geometric} coordinates $q^\mu(x^\mu)$ in the action and by promoting ordinary derivatives to $q$-derivatives, while observables are naturally discussed in the \emph{fractional} (physical) coordinates $x^\mu$.

The multi-scale map $q^\mu(x^\mu)$ depends on a characteristic length scale $\ell_*$ and on fractional exponents, and it may also include logarithmic oscillations. Such oscillations are tied to discrete scale invariance, reviewed by Sornette \cite{Sornette:1997pb}. Log-periodic fractal functions were constructed by Gluzman and Sornette \cite{GluzmanSornette:2002logperiodic}. In the multi-fractional setting, their interpretation in terms of complex dimensions and observability was discussed by Calcagni \cite{Calcagni:2017via}. The same framework is subject to nontrivial phenomenological bounds: particle-physics constraints on multifractal spacetimes were derived in Ref. \cite{Calcagni:2015mxx}, and Standard-Model constraints in multiscale theories were analyzed in Ref. \cite{Calcagni:2015xcf}. Lorentz-violation constraints in multifractal spacetimes were studied in Ref. \cite{Calcagni:2016zqv}. Gravitational-wave luminosity-distance effects in quantum-gravity scenarios were analyzed in Ref. \cite{Calcagni:2019kzo}. Broader multi-messenger quantum-gravity phenomenology is reviewed in Ref. \cite{Addazi:2021xuf}, while recent cosmological tensions and related fundamental-physics issues are surveyed in Ref. \cite{CosmoVerseNetwork:2025alb}. Multi-scale gravity and cosmology were developed in Ref. \cite{Calcagni:2013yqa}. Inflationary and CMB signatures in multi-fractional spacetimes were studied in Ref. \cite{Calcagni:2016ofu}. Related field-theoretic analyses show that effective couplings can inherit scale dependence from the multi-scale structure \cite{Calcagni:2013qqa}. A concrete realization involving a varying electric charge in multiscale spacetimes was given in Ref. \cite{Calcagni:2013yra}, which motivates treating profile parameters as external inputs in variational statements.

Static and radially symmetric black holes in multi-fractional gravity were first analyzed systematically in Ref. \cite{Calcagni:2017ymp}. In the $q$-derivative scenario, the Schwarzschild solution retains its standard form in geometric coordinates, while its physical interpretation in fractional coordinates exhibits genuinely multi-scale effects: the horizon position becomes presentation dependent, and in the final-point presentation an additional (ring-like) singular locus may appear at a radius controlled by $\ell_*$. Ref. \cite{Calcagni:2017ymp} also discusses black-hole temperature in the $q$-derivative theory. Since the theory is equivalent to the ordinary Einstein--Hilbert theory in the geometric coordinates $q^\mu$, the standard geometric-frame Hawking temperature of the Schwarzschild solution is $T_q=(4\pi r_0)^{-1}$. However, physical observables in the multi-fractional interpretation are expressed in the fractional coordinates $x^\mu$. This motivates the fractional-frame temperature prescription used in Ref. \cite{Calcagni:2017ymp}, in which the redshift factor is differentiated with respect to the physical radial coordinate $r$. The resulting quantity differs from $T_q$ by the local measure factor $q'(r_{\rm h})$ and is the temperature prescription analyzed in the present work. More recently, observationally oriented studies have begun exploring potential signatures of multi-fractional corrections in lensing and black-hole shadows \cite{Pantig:2024fbh}.

The purpose of the present paper is to place multi-fractional black-hole thermodynamics on an action-based footing in the static and spherically symmetric vacuum sector of $T_q$, and to reconcile canonical charges with the physical notion of temperature. Concretely, we: (i) formulate the $q$-derivative Einstein--Hilbert action with a well-posed variational principle in the presence of boundaries; (ii) construct the covariant phase space using the Lee--Wald treatment of local symmetries and constraints \cite{Lee:1990nz}, implement the Iyer--Wald Hamiltonian-charge formalism \cite{Iyer:1994ys}, and use Wald's Noether-charge interpretation of black-hole entropy \cite{Wald:1993nt}; (iii) keep track of the boundary structure in a way consistent with the gravitational Hamiltonian analysis of Hawking and Horowitz \cite{Hawking:1995fd}; (iv) evaluate the Hamiltonian charge at infinity and the Noether charge at the horizon for the multi-fractional Schwarzschild family; and (v) derive an \emph{integrable} first law when the Hawking temperature is defined in physical coordinates as in Ref. \cite{Calcagni:2017ymp}. We also clarify the relation between our enlarged state-space construction and the extended Iyer--Wald approach to black-hole thermodynamics. The analogy lies in allowing external parameters to vary and in introducing their conjugate work terms; the difference is that the multi-fractional parameters varied here specify a prescribed geometric-coordinate profile rather than coupling constants in the geometric-frame gravitational Lagrangian. The last step requires extending the thermodynamic state space to include variations of the non-dynamical multi-scale parameters entering $q(r)$, which induces additional, computable work terms conjugate to these parameters. This construction clarifies how presentation dependence and the stochastic interpretation proposed in Ref. \cite{Calcagni:2017ymp} enter black-hole thermodynamics: in the deterministic view, different presentations correspond to inequivalent models with different response functions, while in the stochastic view they delimit intrinsic fluctuations of horizon quantities and thermodynamic variables.

We program the paper as follows. In Sec. \ref{sec2} we summarize the $q$-derivative setup and the static Schwarzschild family written in the geometric radius $q$, together with the branch conditions required for a well-defined exterior region. In Sec. \ref{sec3} we introduce the fractional-frame temperature and frame the associated integrability issue when profile parameters are varied. In Sec. \ref{sec4} we define the canonical mass and the Noether entropy using covariant charge methods. In Sec. \ref{sec5} we formulate an extended first law on the enlarged state space and define an integrable thermodynamic entropy appropriate to the fractional-frame temperature. In Sec. \ref{sec6} we discuss presentation dependence and extend the thermodynamic description to oscillatory multi-fractional measures. We conclude in Sec. \ref{sec7} with consistency conditions and a brief outlook. Throughout, we use metric signature $(-,+,+,+)$ and set $c=\hbar=k_{\rm B}=1$, keeping $G$ explicit.

\section{Multifractional setup and Schwarzschild family in the geometric frame} \label{sec2}
We work in the multi-fractional theory with $q$-derivatives, where the multi-scale structure is encoded in a set of monotonic \emph{geometric coordinates} $q^\mu$ that depend separately on each coordinate $x^\mu$ and are fixed once and for all by the choice of theory. In the \emph{fractional frame} $x^\mu$, fields are evaluated as usual but derivatives are replaced by derivatives with respect to $q^\mu$; in the \emph{geometric frame} $q^\mu$, the theory takes the same functional form as the corresponding integer-dimensional model. In the factorizable setting adopted here, the basic definitions follow the original multi-scale measure construction \cite{Calcagni:2009kc}, its field-theoretic extension \cite{Calcagni:2010bj}, the geometric formulation of multi-fractional spacetime \cite{Calcagni:2011sz}, the detailed geometry of fractional spaces \cite{Calcagni:2011kn}, and the black-hole implementation in the $q$-derivative theory \cite{Calcagni:2017ymp}
\begin{align}
&q^\mu=q^\mu(x^\mu), \nn \\
&v_\mu(x^\mu)=\frac{d q^\mu}{d x^\mu}, \nn \\
&\partial_{q^\mu}=\frac{1}{v_\mu(x^\mu)}\,\partial_\mu,
\label{2.1}
\end{align}
with no sum over $\mu$. The associated Laplacians and momentum-space transforms compatible with fractional measures are discussed in Ref. \cite{Calcagni:2012zj}.
The metric $g_{\mu\nu}$ is a tensor field on the manifold $\mathcal M$, and all geometric quantities in the $q$-derivative theory are obtained from the standard Levi-Civita construction by replacing ordinary derivatives with $\partial_{q^\mu}$. In particular, the torsionless and metric-compatible connection is
\begin{equation}
{}^q\Gamma^{\rho}{}_{\mu\nu}
=\frac12\,g^{\rho\sigma}\Big(\partial_{q^\mu}g_{\nu\sigma}+\partial_{q^\nu}g_{\mu\sigma}-\partial_{q^\sigma}g_{\mu\nu}\Big),
\label{2.2}
\end{equation}
and the corresponding Riemann tensor is \cite{Calcagni:2017ymp}
\begin{equation}
{}^q R^{\rho}{}_{\mu\sigma\nu}
=\partial_{q^\sigma}\,{}^q\Gamma^{\rho}{}_{\mu\nu}
-\partial_{q^\nu}\,{}^q\Gamma^{\rho}{}_{\mu\sigma}
+{}^q\Gamma^{\rho}{}_{\lambda\sigma}\,{}^q\Gamma^{\lambda}{}_{\mu\nu}
-{}^q\Gamma^{\rho}{}_{\lambda\nu}\,{}^q\Gamma^{\lambda}{}_{\mu\sigma}.
\label{2.3}
\end{equation}
Contractions yield ${}^q R_{\mu\nu}={}^q R^{\rho}{}_{\mu\rho\nu}$ and ${}^q R=g^{\mu\nu}{}^q R_{\mu\nu}$, and we define the $q$-Einstein tensor ${}^q G_{\mu\nu}={}^q R_{\mu\nu}-\frac12 g_{\mu\nu}\,{}^q R$.

The gravitational action in $D$ topological dimensions is obtained by the substitution $x^\mu\to q^\mu(x^\mu)$ in the Einstein--Hilbert functional. In the fractional frame it reads
\begin{align}
&{}^q S_{\rm g}[g]
=\frac{1}{2\kappa^2}\int_{\mathcal M} d^D x\; v(x)\,\sqrt{-g}\,\Big({}^q R-2\Lambda\Big),
\nn \\
&v(x)=\prod_{\mu=0}^{D-1}v_\mu(x^\mu),
\label{2.4}
\end{align}
where $\kappa^2$ is the gravitational coupling and $\Lambda$ is the cosmological constant. In the geometric frame, one can regard $q^\mu$ as coordinates and rewrite Eq. \eqref{2.4} as an ordinary Einstein--Hilbert action integrated with $d^D q$, making manifest that the variational structure is the same as in general relativity, up to the replacement of ordinary derivatives by $q$-derivatives \cite{Calcagni:2017ymp}. On manifolds with boundary, the Einstein--Hilbert action does not by itself yield a well-posed Dirichlet problem for the induced metric; in the geometric frame the standard cure is the Gibbons--Hawking--York counterterm introduced in Ref. \cite{Gibbons:1976ue}. The corresponding Hamiltonian and surface-term interpretation is reviewed in Ref. \cite{Hawking:1995fd}. In the present factorizable $q$-theory, the corresponding well-posed action in the fractional frame is obtained by mapping that boundary term back to $x^\mu$ \cite{Calcagni:2017ymp}. For a boundary component $\partial\mathcal M$ with unit normal $n_\mu$ satisfying $n^\mu n_\mu=\epsilon$ with $\epsilon=\pm 1$, we define
\begin{align}
&h_{\mu\nu}=g_{\mu\nu}-\epsilon\,n_\mu n_\nu,\nn \\
&{}^q K_{\mu\nu}=h_\mu{}^{\rho}h_\nu{}^{\sigma}\,{}^q\nabla_{\rho}n_{\sigma},\nn \\
&{}^q K=h^{\mu\nu}\,{}^q K_{\mu\nu},
\label{2.5}
\end{align}
where ${}^q\nabla$ is the covariant derivative built from ${}^q\Gamma^\rho{}_{\mu\nu}$. If the boundary is described by fixing one coordinate $x^\perp={\rm const}$ and the induced coordinates on $\partial\mathcal M$ are $x^a$ ($a=1,\dots,D-1$), then the Jacobian on the boundary is the product of the tangential weights,
\begin{equation}
v_\partial(x)=\prod_{a=1}^{D-1} v_a(x^a)=\frac{v(x)}{v_\perp(x^\perp)}.
\label{2.6}
\end{equation}
The total gravitational action admitting a Dirichlet variational principle for $h_{\mu\nu}$ is therefore
\begin{equation}
{}^q S_{\rm g,tot}[g]
=\frac{1}{2\kappa^2}\int_{\mathcal M} d^D x\; v(x)\,\sqrt{-g}\,\Big({}^q R-2\Lambda\Big)
+\frac{\epsilon}{\kappa^2}\int_{\partial\mathcal M} d^{D-1}x\; v_\partial(x)\,\sqrt{|h|}\;{}^q K,
\label{2.7}
\end{equation}
which is the direct $q$-deformation of the standard Einstein--Hilbert plus Gibbons--Hawking--York functional \cite{Gibbons:1976ue}, with the boundary-Hamiltonian normalization chosen consistently with Ref. \cite{Hawking:1995fd}, and written in a form adapted to factorizable multi-fractional profiles \cite{Calcagni:2017ymp}. Varying the full action ${}^q S_{\rm tot}={}^q S_{\rm g,tot}+{}^q S_{\rm m}$ with respect to $g^{\mu\nu}$ at fixed $q^\mu(x^\mu)$ and fixed induced metric $h_{\mu\nu}$ on $\partial\mathcal M$, the boundary contributions cancel and one obtains the bulk field equations
\begin{equation}
{}^q G_{\mu\nu}+\Lambda g_{\mu\nu}=\kappa^2\,{}^q T_{\mu\nu}.
\label{2.8}
\end{equation}
In the remainder of this work we restrict to vacuum solutions ${}^q T_{\mu\nu}=0$ and, for definiteness, to asymptotically flat configurations with $\Lambda=0$.

For the static and spherically symmetric sector it is convenient to employ the \emph{spherical-coordinate approximation} of Ref. \cite{Calcagni:2017ymp}, where the multi-fractional deformation is implemented through a non-trivial radial profile $q(r)$ while leaving the angular coordinates undeformed. The simplest (binomial) profile in the radial direction is \cite{Calcagni:2017ymp}
\begin{equation}
q(r)=r\pm \frac{\ell_*}{\alpha}\left(\frac{r}{\ell_*}\right)^{\alpha},
\qquad 0<\alpha<1,\qquad r\ge 0,
\label{2.9}
\end{equation}
where $\ell_*$ is the characteristic multi-fractional length scale and the sign encodes a choice of presentation. When logarithmic oscillations are included, the profile generalizes to
\begin{align}
&q(r)=r\pm \frac{\ell_*}{\alpha}\left(\frac{r}{\ell_*}\right)^{\alpha} F_\omega(r),
\nn \\
&F_\omega(r)=1+A\cos\!\left[\omega\ln\!\left(\frac{r}{\ell_\infty}\right)\right]+B\sin\!\left[\omega\ln\!\left(\frac{r}{\ell_\infty}\right)\right],
\label{2.10}
\end{align}
where $\omega$ is a dimensionless log-frequency, $A$ and $B$ are dimensionless amplitudes, and $\ell_\infty$ is an additional length scale associated with discrete scale invariance. The general connection with discrete scale invariance is reviewed in Ref. \cite{Sornette:1997pb}; log-periodic fractal functions are discussed in Ref. \cite{GluzmanSornette:2002logperiodic}; and the role of complex dimensions in the multi-fractional setting is analyzed in Ref. \cite{Calcagni:2017via}. The radial weight function needed to evaluate $q$-derivatives is $v_r(r)=dq/dr$; for the binomial profile (i.e., $F_\omega\equiv 1$) this reduces to
\begin{equation}
v_r(r)=\frac{dq(r)}{dr}=1\pm\left(\frac{r}{\ell_*}\right)^{\alpha-1}.
\label{2.11}
\end{equation}

In the $q$-derivative theory, general-relativistic solutions written in the geometric coordinates $q^\mu(x^\mu)$ solve the $q$-Einstein equations \eqref{2.8} in the fractional frame, once the substitution rule $x^\mu\mapsto q^\mu(x^\mu)$ is implemented in derivatives and in the measure \cite{Calcagni:2017ymp}. In particular, in the spherical-coordinate approximation where only the radial direction is deformed and $(t,\theta,\phi)$ are left undeformed, the vacuum static and spherically symmetric solution can be written directly by importing the Schwarzschild form in the areal coordinate $q$ \cite{Calcagni:2017ymp}:
\begin{equation}
{}^{q}\!ds^{2}
=
-\left(1-\frac{r_{0}}{q}\right)dt^{2}
+\left(1-\frac{r_{0}}{q}\right)^{-1}dq^{2}
+q^{2}\,d\Omega_{2}^{2},
\label{2.12}
\end{equation}
where $d\Omega_{2}^{2}=d\theta^{2}+\sin^{2}\theta\,d\phi^{2}$. Here $r_{0}$ is an integration constant that reduces to the usual Schwarzschild radius in the general-relativistic limit; following Ref. \cite{Calcagni:2017ymp} we keep the conventional notation $r_{0}=2GM$ for later comparison, while postponing the action-based identification of the conserved mass to the covariant phase-space/Hamiltonian analysis.

In the fractional frame, the geometric radius is a prescribed function $q=q(r)$ of the fractional radial coordinate $r$, given by Eq. \eqref{2.9} or Eq. \eqref{2.10}. Substituting $q=q(r)$ into Eq. \eqref{2.12} yields a line element on the coordinate chart $(t,r,\theta,\phi)$ of the form
\begin{equation}
{}^{q}\!ds^{2}
=
-f(r)\,dt^{2}
+f(r)^{-1}\left(\frac{dq(r)}{dr}\right)^{2}dr^{2}
+q^{2}(r)\,d\Omega_{2}^{2},
\label{2.13}
\end{equation}
where we introduced the shorthand
\begin{equation}
f(r)=1-\frac{r_{0}}{q(r)}.
\label{2.14}
\end{equation}
The normalization of the static Killing field $\partial_{t}$ is fixed by requiring $g_{tt}\to-1$ as $r\to\infty$, which is consistent with the admissible asymptotics $q(r)\sim r$ for $0<\alpha<1$ and bounded oscillatory modulation, so that $f(r)\to 1$ at large radii.

The event horizon, when it exists, is a Killing horizon of $\partial_t$ and is located by the condition $g_{tt}=0$. Using Eq. \eqref{2.14}, this gives the implicit horizon equation
\begin{equation}
q(r_{\rm h})=r_{0}.
\label{2.15}
\end{equation}
For the coarse-grained binomial profile \eqref{2.9} without logarithmic oscillations, Eq. \eqref{2.15} reduces to the explicit algebraic relation
\begin{equation}
r_{\rm h}\,\pm\,\frac{\ell_*^{\,1-\alpha}}{\alpha}\,r_{\rm h}^{\alpha}=r_{0}.
\label{2.16}
\end{equation}

Since the coefficients in Eqs. \eqref{2.13} and \eqref{2.14} depend explicitly on $q(r)$, zeros of the geometric radius mark loci where metric components diverge and the standard Schwarzschild curvature singularity in the areal coordinate $q$ is inherited by the multi-fractional solution. In particular, for the binomial profile in the final-point presentation one has, besides $q(0)=0$, an additional zero at finite $r$ \cite{Calcagni:2017ymp}:
\begin{equation}
q(r)=r-\frac{\ell_*^{\,1-\alpha}}{\alpha}\,r^\alpha=0
\qquad\Longleftrightarrow\qquad
r=\alpha^{-\frac{1}{1-\alpha}}\ell_* \sim \ell_* .
\label{2.17}
\end{equation}
In the present work, whenever we discuss black-hole exteriors and associated charges, we restrict to an outer branch satisfying the following conditions: (i) $q(r)>0$ for all $r\ge r_{\rm h}$, so that the areal radius is real and the angular part is non-degenerate; (ii) $q'(r)>0$ for all $r\ge r_{\rm h}$, so that $q$ is a good radial coordinate in the exterior; and (iii) Eq. \eqref{2.15} admits a unique solution $r_{\rm h}$ on that exterior branch, so that horizon quantities are single-valued functions of the parameters. With logarithmic oscillations, Eq. \eqref{2.10}, additional ultra-short-distance zeros can occur when amplitudes are near maximal values \cite{Calcagni:2017ymp}; throughout we treat oscillations as controlled deformations and impose the same exterior regularity requirements.

\section{Physical temperature in fractional coordinates and the integrability issue} \label{sec3}
In the static vacuum geometry introduced in Sec. \ref{sec2}, the timelike Killing vector is $\xi=\partial_t$, normalized by the asymptotic condition $g_{tt}\to-1$ as $r\to\infty$. Its norm is
\begin{equation}
\xi^{2}=g_{\mu\nu}\xi^\mu\xi^\nu=g_{tt}=-f(r),
\label{3.1}
\end{equation}
with $f(r)$ given in Eq. \eqref{2.14}. A (non-extremal) Killing horizon is located at the largest root $r=r_{\rm h}$ of $f(r)=0$, equivalently by the horizon condition $q(r_{\rm h})=r_{0}$ in Eq. \eqref{2.15}. Throughout this section we remain on an exterior branch where $q(r)$ is positive and monotonic in a neighborhood of $r_{\rm h}$, so that the exterior region is well defined and $q'(r_{\rm h})$ exists.

It is useful to distinguish two temperature notions. In the geometric frame, where the solution is the ordinary Schwarzschild metric written in the areal coordinate $q$, the standard semiclassical Euclidean argument gives
\begin{equation}
T_q=\frac{1}{4\pi r_0}.
\label{3.2a}
\end{equation}
This is the temperature associated with the regularity of the Euclidean section in the geometric coordinate. In the multi-fractional interpretation, however, observables are described in the fractional coordinates. Following the prescription adopted in Ref. \cite{Calcagni:2017ymp}, we therefore define the corresponding fractional-frame temperature by differentiating the redshift factor with respect to the physical radial coordinate $r$:
\begin{equation}
T_{\rm h}=\frac{1}{4\pi}\,\frac{d}{dr}\left[1-\frac{r_{0}}{q(r)}\right]\Bigg|_{r=r_{\rm h}}.
\label{3.2}
\end{equation}
In what follows, $T_{\rm h}$ denotes this fractional-frame temperature, not the geometric-frame temperature $T_q$. It is convenient to introduce the associated surface-gravity parameter $\kappa_{\rm h}$ via
\begin{equation}
\kappa_{\rm h}=2\pi\,T_{\rm h},
\label{3.3}
\end{equation}
so that Eq. \eqref{3.2} is equivalently $\kappa_{\rm h}=\frac12 f'(r_{\rm h})$ with the derivative taken with respect to $r$. Using $f(r)=1-r_0/q(r)$, we obtain
\begin{align}
&\kappa_{\rm h}
=\frac12\,\frac{d}{dr}\left(1-\frac{r_{0}}{q(r)}\right)\Bigg|_{r=r_{\rm h}}
=\frac{r_{0}}{2}\,\frac{q'(r_{\rm h})}{q^{2}(r_{\rm h})},
\nn \\
&q'(r)=\frac{dq(r)}{dr},
\label{3.4}
\end{align}
and, upon imposing $q(r_{\rm h})=r_{0}$,
\begin{align}
&\kappa_{\rm h}=\frac{q'(r_{\rm h})}{2r_{0}},
\nn \\
&T_{\rm h}=\frac{q'(r_{\rm h})}{4\pi r_{0}}.
\label{3.5}
\end{align}
Therefore, within the fractional-frame definition \eqref{3.2}, deviations of the evaporation temperature from the canonical Schwarzschild value are controlled entirely by the local weight factor $q'(r_{\rm h})$ at the horizon.

For the binomial profile without logarithmic oscillations, Eq. \eqref{2.9}, one has $q'(r)=v_r(r)$ given in Eq. \eqref{2.11}, and Eq. \eqref{3.5} yields
\begin{equation}
T_{\rm h}
=\frac{1}{4\pi r_{0}}
\left[1\pm\left(\frac{r_{\rm h}}{\ell_*}\right)^{\alpha-1}\right],
\qquad (0<\alpha<1),
\label{3.6}
\end{equation}
where the sign is the same as in the definition of $q(r)$ and $r_{\rm h}$ is determined implicitly by Eq. \eqref{2.16}. Since $\alpha-1<0$, the correction in brackets is parametrically suppressed for large horizons $r_{\rm h}\gg \ell_*$, consistently with the expectation that multi-fractional effects are negligible in the infrared. When logarithmic oscillations are present, Eq. \eqref{2.10}, the relation $T_{\rm h}=q'(r_{\rm h})/(4\pi r_0)$ remains valid, but $q'(r)$ acquires an oscillatory modulation. Writing Eq. \eqref{2.10} as $q(r)=r\pm (\ell_*^{\,1-\alpha}/\alpha)\,r^{\alpha}F_\omega(r)$, direct differentiation gives
\begin{align}
&q'(r)=1\pm \ell_*^{\,1-\alpha}\,r^{\alpha-1}F_\omega(r)\pm \frac{\ell_*^{\,1-\alpha}}{\alpha}\,r^{\alpha}F_\omega'(r),
\nn \\
&F_\omega'(r)=\frac{\omega}{r}\Bigg[-A\sin\!\Bigg(\omega\ln\frac{r}{\ell_\infty}\Bigg)+B\cos\!\Bigg(\omega\ln\frac{r}{\ell_\infty}\Bigg)\Bigg],
\label{3.7}
\end{align}
so that $T_{\rm h}$ inherits oscillations as a function of $r_{\rm h}$.
\begin{figure}
    \centering
    \includegraphics[width=0.6\linewidth]{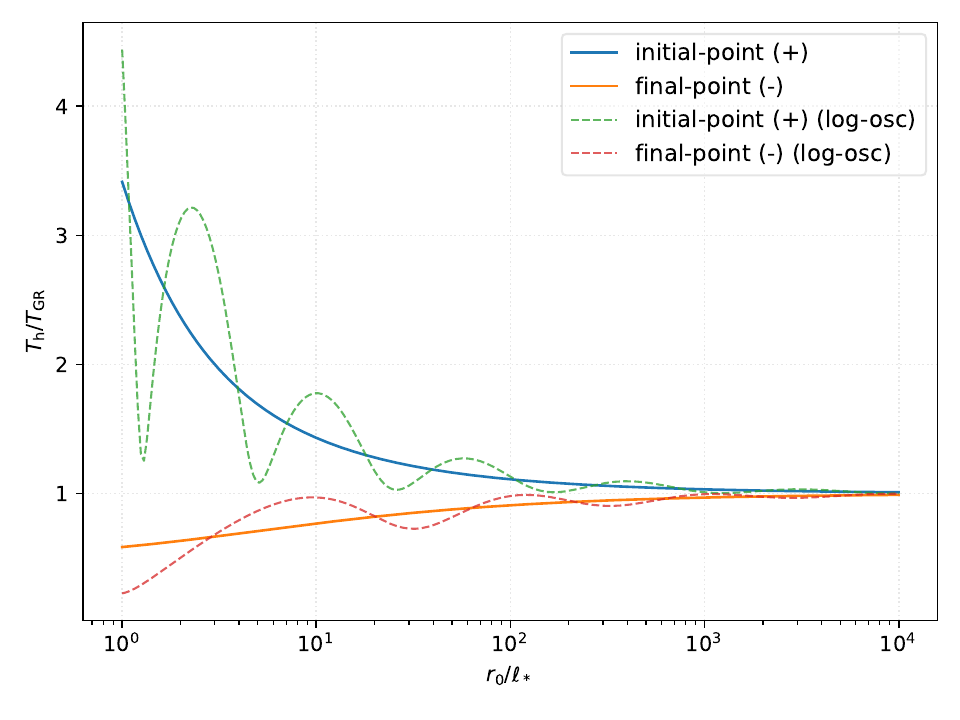}
    \caption{Temperature ratio $T_{\rm h}/T_{\rm GR}=q'(r_{\rm h})$ versus $r_0/\ell_*$ for the binomial profile (solid curves) in the initial-point and final-point presentations. The dashed curves show a representative log-oscillating profile with small amplitudes. Here, $\alpha = 0.50$, and $\ell_* = 1.0$.}
    \label{fig1}
\end{figure}
Figure \ref{fig1} visualizes the key distinction between canonical and thermal quantities in the fractional frame. For the binomial profile, the ratio $T_{\rm h}/T_{\rm GR}=q'(r_{\rm h})$ departs from unity by a correction controlled by $(r_{\rm h}/\ell_*)^{\alpha-1}$, and therefore approaches the general-relativistic value in the infrared. When logarithmic oscillations are included, the same ratio acquires a bounded log-periodic modulation, as expected from Eq. \eqref{3.7}; in the illustrative curves we used $A=0.12$, $B=0.08$, $\omega=3$, and $\ell_\infty=\ell_*$. Throughout, the plotted branches are restricted to configurations with $q'(r_{\rm h})>0$ so that $T_{\rm h}$ remains positive and the exterior branch is well defined.

Let us also clarify the relation with the usual Euclidean regularity argument. Applying the conical-regularity calculation to the pulled-back metric \eqref{2.13} gives
\begin{equation}
ds_E^{2}\simeq f'(r_{\rm h})(r-r_{\rm h})\,d\tau^{2}
+\frac{q'(r_{\rm h})^{2}}{f'(r_{\rm h})(r-r_{\rm h})}\,dr^{2},
\label{3.8}
\end{equation}
near the horizon. With the proper radial coordinate
\begin{equation}
\rho=\frac{2q'(r_{\rm h})}{\sqrt{f'(r_{\rm h})}}\sqrt{r-r_{\rm h}},
\label{3.9}
\end{equation}
one obtains
\begin{equation}
ds_E^{2}\simeq d\rho^{2}
+\left(\frac{f'(r_{\rm h})}{2q'(r_{\rm h})}\right)^{2}\rho^{2}\,d\tau^{2}.
\label{3.10}
\end{equation}
Regularity therefore requires the Euclidean period
\begin{equation}
\beta_q=\frac{4\pi q'(r_{\rm h})}{f'(r_{\rm h})}.
\label{3.11}
\end{equation}
Since $f'(r_{\rm h})=q'(r_{\rm h})/r_0$, this gives $\beta_q=4\pi r_0$ and hence the geometric-frame Hawking temperature $T_q=(4\pi r_0)^{-1}$. Thus the factor $q'(r_{\rm h})$ cancels in the standard conical-regularity calculation. The fractional-frame quantity in Eq. \eqref{3.2} should therefore be regarded as an operational temperature prescription tied to differentiation with respect to the physical radial coordinate $r$, rather than as an independent conical-regularity temperature. The distinction between $T_q$ and $T_{\rm h}=q'(r_{\rm h})T_q$ is central to the thermodynamic construction below.

If one instead uses the geometric-frame temperature $T_q$, the canonical Iyer--Wald first law retains its standard Schwarzschild form with the Noether entropy $S_{\rm N}$. The nontrivial integrability issue analyzed below arises only after adopting the fractional-frame temperature prescription \eqref{3.2}.

We now explain why the temperature definition \eqref{3.2} leads to a nontrivial integrability issue once the multifractional profile is allowed to vary. Let $\lambda^I$ denote the set of non-dynamical parameters entering $q(r)$ (for instance $\ell_*,\alpha$ and, when present, $A,B,\omega,\ell_\infty$ as in Ref. \cite{Calcagni:2017ymp}). Then $T_{\rm h}$ depends on both the integration constant $r_0$ and the profile parameters through the horizon location determined by $q(r_{\rm h},\lambda)=r_0$ and through the local factor $q'(r_{\rm h},\lambda)$ in Eq. \eqref{3.5}. By contrast, the canonical charges of the solution are defined covariantly in the geometric frame and are controlled by the asymptotic and horizon data of the Schwarzschild family written in the areal coordinate $q$; in particular, for fixed $\lambda^I$ the one-parameter family labeled by $r_0$ admits an integrable first law.

The difficulty arises if one insists on using $T_{\rm h}$ as the physical temperature entering a Clausius relation while simultaneously allowing variations of the profile parameters. To see the obstruction at the level of differential forms, consider the 1-form on the extended parameter space spanned by $(r_0,\lambda^I)$,
\begin{equation}
\vartheta=\frac{1}{T_{\rm h}(r_0,\lambda)}\,dM(r_0),
\label{3.12}
\end{equation}
where $M(r_0)$ denotes the conserved mass regarded as a function of the Schwarzschild parameter $r_0$ (its explicit form will be fixed by the canonical charge at infinity in the next section). If $\vartheta$ were the exact differential of an entropy state function on the extended space, then $d\vartheta$ would vanish identically. However, since $dM$ has no components along $d\lambda^I$ while $T_{\rm h}$ generically does, one finds
\begin{equation}
d\vartheta
=-\frac{dM(r_0)}{T_{\rm h}(r_0,\lambda)^{2}}\wedge dT_{\rm h}(r_0,\lambda),
\label{3.13}
\end{equation}
which is nonzero whenever $\partial_{\lambda^I}T_{\rm h}\neq 0$. Therefore, unless the profile is held fixed, the naive Clausius identification $dS=dM/T_{\rm h}$ does not, in general, define an integrable thermodynamic entropy on the enlarged space. The resolution is to enlarge the first law by admitting additional work terms conjugate to the variations $d\lambda^I$, thereby restoring exactness of the energy differential on the extended state space. This will be implemented explicitly once the canonical charges are in hand.

\section{Canonical charges: mass and Noether entropy} \label{sec4}
In the $q$-derivative theory, the gravitational dynamics is Einstein--Hilbert in the geometric frame, and the conserved charges associated with diffeomorphisms can be extracted efficiently with the covariant phase-space construction of Lee, Wald, and Iyer \cite{Iyer:1994ys}. Related covariant constructions of conserved quantities and their fluxes at boundaries include the Wald--Zoupas prescription \cite{Wald:1999wa}. A cohomological treatment of asymptotic symmetries, conservation laws, and central charges was developed by Barnich and Brandt \cite{Barnich:2001jy}. Here we employ the Iyer--Wald framework tailored to the stationary sector. We briefly recall only the identities needed for the stationary family at hand. Let $\phi$ denote the dynamical fields, which in our vacuum application reduce to $\phi\equiv g_{ab}$. The Lagrangian is a spacetime 4-form $\bm L(\phi)$, and its first variation admits the universal decomposition
\begin{equation}
\delta \bm L=\bm E\,\delta\phi+d\bm\Theta(\phi,\delta\phi),
\label{4.1}
\end{equation}
where $\bm E=0$ encodes the equations of motion and $\bm\Theta(\phi,\delta\phi)$ is the symplectic potential 3-form \cite{Iyer:1994ys}. For four-dimensional vacuum general relativity, which is the relevant geometric-frame dynamics here, one may take
\begin{align}
&\bm L_{abcd}=\frac{1}{2\kappa^2}\,\epsilon_{abcd}\,R,
\nn \\
&\kappa^2=8\pi G,
\label{4.2}
\end{align}
together with the associated symplectic potential \cite{Iyer:1994ys}
\begin{equation}
\bm\Theta_{abc}
=\epsilon_{dabc}\,\frac{1}{2\kappa^2}\,g^{de}g^{fh}\big(\nabla_f\delta g_{eh}-\nabla_e\delta g_{fh}\big),
\label{4.3}
\end{equation}
where $\nabla$ is the Levi--Civita connection of $g_{ab}$ and $\epsilon_{abcd}$ is the metric volume form.

Given a vector field $\xi^a$ generating an infinitesimal diffeomorphism, the Noether current 3-form is defined as \cite{Iyer:1994ys}
\begin{equation}
\bm J[\xi]=\bm\Theta(\phi,\mathcal{L}_\xi\phi)-\xi\cdot\bm L,
\label{4.4}
\end{equation}
where $\xi\cdot\bm L$ denotes interior contraction of $\xi^a$ into the first index of $\bm L$. Diffeomorphism covariance implies $d\bm J[\xi]=-\bm E\,\mathcal{L}_\xi\phi$, so that on shell ($\bm E=0$) the current is closed and can be written as
\begin{equation}
\bm J[\xi]=d\bm Q[\xi]\qquad(\bm E=0),
\label{4.5}
\end{equation}
defining the Noether charge 2-form $\bm Q[\xi]$ \cite{Iyer:1994ys}. For vacuum general relativity, the explicit charge 2-form is \cite{Iyer:1994ys}
\begin{equation}
\bm Q_{ab}[\xi]=-\frac{1}{2\kappa^2}\,\epsilon_{abcd}\,\nabla^{c}\xi^{d}.
\label{4.6}
\end{equation}
If $\xi^a$ is a symmetry of a background solution and $\delta g_{ab}$ satisfies the linearized equations about that background, the fundamental variational identity is the closure of the 2-form $(\delta\bm Q[\xi]-\xi\cdot\bm\Theta)$, so that for any hypersurface $\Xi$ one has \cite{Iyer:1994ys}
\begin{equation}
\int_{\partial\Xi}\Big(\delta\bm Q[\xi]-\xi\cdot\bm\Theta(\phi,\delta\phi)\Big)=0.
\label{4.7}
\end{equation}
For our purposes, $\Xi$ may be chosen to connect a horizon cross-section to a large sphere at spatial infinity, with $\xi=\partial_t$. The term at infinity is interpreted as the variation of the Hamiltonian generator of the asymptotic symmetry, i.e.\ the canonical energy (mass) \cite{Iyer:1994ys}.

To compute the mass explicitly for the multifractional Schwarzschild family, we use the fact that in the geometric frame the line element is exactly Schwarzschild in the areal coordinate $q$,
\begin{align}
&ds^{2}=-f(q)\,dt^{2}+f(q)^{-1}dq^{2}+q^{2}d\Omega_{2}^{2},
\nn \\
&f(q)=1-\frac{r_{0}}{q},
\label{4.8}
\end{align}
with asymptotic normalization fixed by $g_{tt}\to-1$. For Einstein gravity with Dirichlet boundary conditions, the canonical energy associated with $\xi=\partial_t$ can be computed as the Hawking--Horowitz Hamiltonian surface charge at infinity with Minkowski subtraction \cite{Hawking:1995fd}. This construction is consistent with the role of surface integrals in the Hamiltonian formulation emphasized by Regge and Teitelboim \cite{Regge:1974zd}, and with the quasilocal energy formulation of Brown and York \cite{Brown:1992br}. Evaluating that surface term for Eq. \eqref{4.8} yields the on-shell energy $E=r_0/2$ in units $G=1$ \cite{Hawking:1995fd}, and restoring Newton's constant gives the canonical mass
\begin{align}
&M\equiv E=\frac{r_{0}}{2G},
\nn \\
&\delta M=\frac{1}{2G}\,\delta r_{0}.
\label{4.9}
\end{align}
For completeness, in backgrounds with nonzero cosmological constant one may define conserved charges relative to the maximally symmetric reference metric using the Abbott--Deser construction. \cite{Abbott:1981ff}. A point that will be used repeatedly is that $M$ is controlled solely by the coefficient of the $1/q$ falloff in $g_{tt}$ in the geometric frame. In the fractional frame, admissible profiles obey $q(r)=r+\mathcal{O}(r^\alpha)$ with $0<\alpha<1$ (and bounded oscillatory modulation when present), implying that variations of multifractional profile parameters affect only subleading asymptotics and therefore do not modify the canonical charge \eqref{4.9} at fixed $r_0$.

We next extract the Noether entropy from the horizon contribution in the identity \eqref{4.7}. For an asymptotically flat stationary black hole with a bifurcate Killing horizon generated by $\xi^a$, Iyer and Wald show that the horizon term can be written as a surface-gravity factor times the variation of an entropy functional constructed from the Noether charge \cite{Iyer:1994ys}. Related physical-process and null-surface formulations of black-hole thermodynamics provide useful context. Mishra, Chakraborty, Ghosh, and Sarkar analyzed the physical-process first law for dynamical black holes \cite{Mishra:2017sqs}. Aneesh, Chakraborty, Hoque, and Virmani extended first-law considerations to systems with fermions \cite{Aneesh:2020fcr}. Chakraborty and Padmanabhan developed a thermodynamic interpretation of geometrical variables associated with null surfaces \cite{Chakraborty:2015hna}. In our static, non-rotating case, the first law reduces to
\begin{equation}
\delta M=\frac{\kappa}{2\pi}\,\delta S_{\rm N},
\label{4.10}
\end{equation}
where $\kappa$ is the surface gravity of the horizon Killing field $\xi=\partial_t$ computed in the geometric frame, and $S_{\rm N}$ is the Iyer--Wald (Noether-charge) entropy \cite{Iyer:1994ys}. Related aspects of the interpretation of black-hole entropy in generally covariant theories were discussed by Jacobson, Kang, and Myers \cite{Jacobson:1993vj}. For the Schwarzschild form \eqref{4.8}, the horizon is at $q=r_0$ and the geometric-frame surface gravity is
\begin{equation}
\kappa=\frac12\,\frac{d}{dq}\!\left(1-\frac{r_0}{q}\right)\Bigg|_{q=r_0}=\frac{1}{2r_0}.
\label{4.11}
\end{equation}
Moreover, for the Einstein--Hilbert Lagrangian, the Iyer--Wald prescription gives the standard area law \cite{Iyer:1994ys}
\begin{equation}
S_{\rm N}=\frac{\mathrm{Area}[\Sigma]}{4G},
\label{4.12}
\end{equation}
where $\Sigma$ is any spatial cross-section of the event horizon (conveniently the bifurcation two-sphere). For Eq. \eqref{4.8}, the induced metric on $\Sigma$ is $d\sigma^2=q^2 d\Omega_2^2|_{q=r_0}=r_0^2 d\Omega_2^2$, hence
\begin{align}
&\mathrm{Area}[\Sigma]=4\pi r_0^2,
\nn \\
&S_{\rm N}=\frac{\pi r_0^2}{G},
\nn \\
&\delta S_{\rm N}=\frac{2\pi r_0}{G}\,\delta r_0.
\label{4.13}
\end{align}
Equations \eqref{4.9}, \eqref{4.11}, and \eqref{4.13} satisfy Eq. \eqref{4.10} identically. Importantly, $S_{\rm N}$ depends only on the geometric horizon radius $q(r_{\rm h})=r_0$ and not on the detailed form of the prescribed map $q=q(r)$: varying multifractional profile parameters at fixed $r_0$ does not change the horizon area in the geometric frame. This profile-insensitivity of the canonical pair $(M,S_{\rm N})$ is precisely what clashes with the profile-sensitive physical temperature $T_{\rm h}$ in Sec. \ref{sec3} once theory parameters are allowed to vary.

\section{Extended first law and an integrable thermodynamic entropy} \label{sec5}
In Sec. \ref{sec4} we found that the canonical mass and the Noether (Iyer--Wald) entropy are fixed solely by the geometric Schwarzschild parameter $r_0$ and satisfy the standard first law with the geometric-frame surface gravity $\kappa$ in Eq. \eqref{4.11}. In Sec. \ref{sec3}, instead, the temperature defined in the fractional radial coordinate $r$ is
\begin{align}
&T_{\rm h}=\frac{q'(r_{\rm h})}{4\pi r_0},
\nn \\
&r_0=q(r_{\rm h}),
\label{5.1}
\end{align}
where $r_{\rm h}$ is the (outer-branch) solution of Eq. \eqref{2.15}. Since $q(r)$ depends on the non-dynamical multi-fractional parameters, the temperature $T_{\rm h}$ generically depends on those parameters even at fixed $r_0$, while $M=r_0/(2G)$ does not. The purpose of this section is to construct a thermodynamic formulation that (i) uses $T_{\rm h}$ as the temperature, (ii) admits variations of the profile parameters, and (iii) remains integrable on an extended state space. The general strategy of enlarging black-hole thermodynamics by allowing variations of external parameters has close analogs in extended black-hole thermodynamics. Kastor, Ray, and Traschen interpreted the black-hole mass as enthalpy when the cosmological constant is varied in AdS black-hole mechanics \cite{Kastor:2009wy}. Dolan developed the pressure--volume interpretation in the first law \cite{Dolan:2011xt}. Kastor, Ray, and Traschen also derived a Smarr formula and extended first law for Lovelock gravity with variable couplings \cite{Kastor:2010gq}. In particular, the extended Iyer--Wald formalism of Xiao, Tian, and Liu \cite{Xiao:2023lap} provides a covariant derivation of such first laws by introducing an extended variation acting both on the dynamical fields and on coupling parameters appearing in a diffeomorphism-invariant gravitational Lagrangian. The present construction is closely related in spirit, since the parameters $\lambda^I$ are also treated as external thermodynamic variables. However, there is an important distinction: in the $q$-derivative Schwarzschild sector considered here, $\lambda^I$ do not play the role of higher-curvature couplings or of a cosmological constant in the geometric-frame Einstein--Hilbert Lagrangian. They are non-dynamical parameters specifying the prescribed multi-fractional profile $q(r,\lambda)$. Therefore, the additional terms derived below should be interpreted as profile-response work terms associated with the fractional-frame temperature prescription, rather than as thermodynamic volumes generated by explicit coupling dependence of the gravitational Lagrangian.

We denote collectively by $\{\lambda^I\}$ the set of parameters appearing in the profile $q(r,\lambda)$, such as $(\ell_*,\alpha)$ for the binomial profile and, when present, $(A,B,\omega,\ell_\infty)$ for logarithmic oscillations \cite{Calcagni:2017ymp}. The horizon is defined implicitly by
\begin{equation}
q(r_{\rm h},\lambda)=r_0,
\label{5.2}
\end{equation}
and therefore, upon varying both $r_0$ and the parameters, we obtain the exact differential identity
\begin{align}
&dr_0=q'(r_{\rm h},\lambda)\,dr_{\rm h}+\Big(\partial_I q\Big)_{\rm h}\,d\lambda^I,
\nn \\
&\Big(\partial_I q\Big)_{\rm h}=\left.\frac{\partial q(r,\lambda)}{\partial\lambda^I}\right|_{r=r_{\rm h}},
\label{5.3}
\end{align}
where $q'(r)=dq/dr$ and repeated $I$ indices are summed. Using the canonical mass in Eq. \eqref{4.9}, the variation of $M$ is
\begin{equation}
dM=\frac{1}{2G}\,dr_0=\frac{1}{2G}\,q'(r_{\rm h},\lambda)\,dr_{\rm h}+\frac{1}{2G}\Big(\partial_I q\Big)_{\rm h}\,d\lambda^I.
\label{5.4}
\end{equation}
The first term in Eq. \eqref{5.4} is the ``horizon-motion'' contribution along $dr_{\rm h}$, while the second term is purely induced by varying the multi-fractional profile at fixed horizon location.

We now define an integrable thermodynamic entropy $S_{\rm th}$ by requiring that, for fixed profile parameters ($d\lambda^I=0$), the Clausius relation with the physical temperature holds:
\begin{equation}
dM=T_{\rm h}\,dS_{\rm th}\qquad (d\lambda^I=0).
\label{5.5}
\end{equation}
Combining Eqs. \eqref{5.1}, \eqref{5.4}, and \eqref{5.5}, we obtain the radial derivative of $S_{\rm th}$ along the family at fixed $\lambda$,
\begin{equation}
\left(\frac{\partial S_{\rm th}}{\partial r_{\rm h}}\right)_{\lambda}
=\frac{1}{T_{\rm h}}\left(\frac{\partial M}{\partial r_{\rm h}}\right)_{\lambda}
=\frac{1}{\frac{q'(r_{\rm h})}{4\pi r_0}}\left(\frac{1}{2G}\,q'(r_{\rm h})\right)
=\frac{2\pi}{G}\,r_0
=\frac{2\pi}{G}\,q(r_{\rm h}),
\label{5.6}
\end{equation}
where in the last step we used Eq. \eqref{5.2}. Integrating Eq. \eqref{5.6} from the origin to the horizon yields a state function (for each fixed presentation and exterior branch)
\begin{equation}
S_{\rm th}(r_{\rm h},\lambda)=\frac{2\pi}{G}\int_{0}^{r_{\rm h}} q(\bar r,\lambda)\,d\bar r.
\label{5.7}
\end{equation}
This choice is natural because it reduces to the Noether entropy when multi-fractional effects are switched off: if $q(r)=r$, then $r_{\rm h}=r_0$ and Eq. \eqref{5.7} gives $S_{\rm th}=\pi r_0^2/G=S_{\rm N}$ from Eq. \eqref{4.13}. For a generic multi-fractional profile, Eq. \eqref{5.7} differs from $S_{\rm N}$ while remaining integrable by construction with respect to variations of the horizon location at fixed $\lambda^I$.

We now allow variations of the profile parameters and write the full differential of $S_{\rm th}$ implied by Eq. \eqref{5.7}:
\begin{equation}
dS_{\rm th}
=\frac{2\pi}{G}\,q(r_{\rm h},\lambda)\,dr_{\rm h}
+\frac{2\pi}{G}\left(\int_{0}^{r_{\rm h}}\partial_I q(\bar r,\lambda)\,d\bar r\right)d\lambda^I.
\label{5.8}
\end{equation}
Multiplying Eq. \eqref{5.8} by $T_{\rm h}$ and using Eq. \eqref{5.1} gives
\begin{equation}
T_{\rm h}\,dS_{\rm th}
=\frac{1}{2G}\,q'(r_{\rm h},\lambda)\,dr_{\rm h}
+\frac{q'(r_{\rm h},\lambda)}{2G\,q(r_{\rm h},\lambda)}
\left(\int_{0}^{r_{\rm h}}\partial_I q(\bar r,\lambda)\,d\bar r\right)d\lambda^I.
\label{5.9}
\end{equation}
Comparing Eq. \eqref{5.9} with the exact variation of the mass in Eq. \eqref{5.4}, we obtain an extended first law of the form
\begin{equation}
dM=T_{\rm h}\,dS_{\rm th}+\Psi_I\,d\lambda^I,
\label{5.10}
\end{equation}
where the conjugate ``multi-fractional potentials'' $\Psi_I$ are determined uniquely by integrability to be
\begin{equation}
\Psi_I
=\frac{1}{2G}\Big(\partial_I q\Big)_{\rm h}
-\frac{q'(r_{\rm h},\lambda)}{2G\,q(r_{\rm h},\lambda)}
\left(\int_{0}^{r_{\rm h}}\partial_I q(\bar r,\lambda)\,d\bar r\right).
\label{5.11}
\end{equation}
Equations \eqref{5.7}, \eqref{5.10}, and \eqref{5.11} resolve the non-integrability noted in Sec. \ref{sec3}: the profile dependence of $T_{\rm h}$ is accounted for by enlarging the thermodynamic state space and by including the work terms $\Psi_I d\lambda^I$. This construction is purely kinematical once the fractional-frame temperature prescription \eqref{3.2} is adopted, and it is compatible with the canonical mass \eqref{4.9} because the $\lambda^I$ are treated as external, non-dynamical parameters specifying the measure profile, as in the black-hole analysis of Ref. \cite{Calcagni:2017ymp}. The same external-parameter viewpoint is compatible with the treatment of varying couplings in multi-scale field theory \cite{Calcagni:2013qqa} and with the varying-charge realization in multiscale spacetimes \cite{Calcagni:2013yra}.

It is useful to compare Eq. \eqref{5.10} with the extended Iyer--Wald first law. In the formalism of Ref. \cite{Xiao:2023lap}, one considers a gravitational Lagrangian whose couplings, such as $\Lambda$ or higher-curvature coefficients, are also varied. The extended variation then produces source terms proportional to the coupling variations, and the corresponding conjugates can be interpreted as generalized thermodynamic volumes. In the present problem, by contrast, the geometric-frame dynamics remains the Einstein--Hilbert dynamics of the Schwarzschild family, and the canonical Iyer--Wald quantities remain
\begin{equation}
M=\frac{r_0}{2G},
\qquad
S_{\rm N}=\frac{\pi r_0^2}{G}.
\end{equation}
The profile parameters $\lambda^I$ enter through the prescribed map $q(r,\lambda)$ and hence through the fractional-frame temperature and horizon location. Thus the potentials $\Psi_I$ in Eq. \eqref{5.11} are not geometric volumes obtained from an explicit $\partial L/\partial \lambda^I$ term in the Lagrangian. Rather, they are the response coefficients required to make the fractional-frame thermodynamic one-form integrable on the enlarged space $(r_{\rm h},\lambda^I)$. If one instead uses the geometric-frame temperature $T_q$ and the Noether entropy $S_{\rm N}$, the ordinary Schwarzschild Iyer--Wald first law is recovered without these additional profile-work terms.

It is useful to display explicit closed forms for the binomial profile of Eq. \eqref{2.9} without logarithmic oscillations. Writing
\begin{equation}
q(r)=r\pm \frac{\ell_*^{\,1-\alpha}}{\alpha}\,r^\alpha,
\qquad 0<\alpha<1,
\label{5.12}
\end{equation}
Eq. \eqref{5.7} gives
\begin{equation}
S_{\rm th}
=\frac{2\pi}{G}\int_{0}^{r_{\rm h}}\left(\bar r\pm \frac{\ell_*^{\,1-\alpha}}{\alpha}\,\bar r^\alpha\right)d\bar r
=\frac{\pi}{G}\,r_{\rm h}^{2}\pm \frac{2\pi}{G}\,\frac{\ell_*^{\,1-\alpha}}{\alpha(\alpha+1)}\,r_{\rm h}^{\alpha+1}.
\label{5.13}
\end{equation}
\begin{figure}
    \centering
    \includegraphics[width=0.6\linewidth]{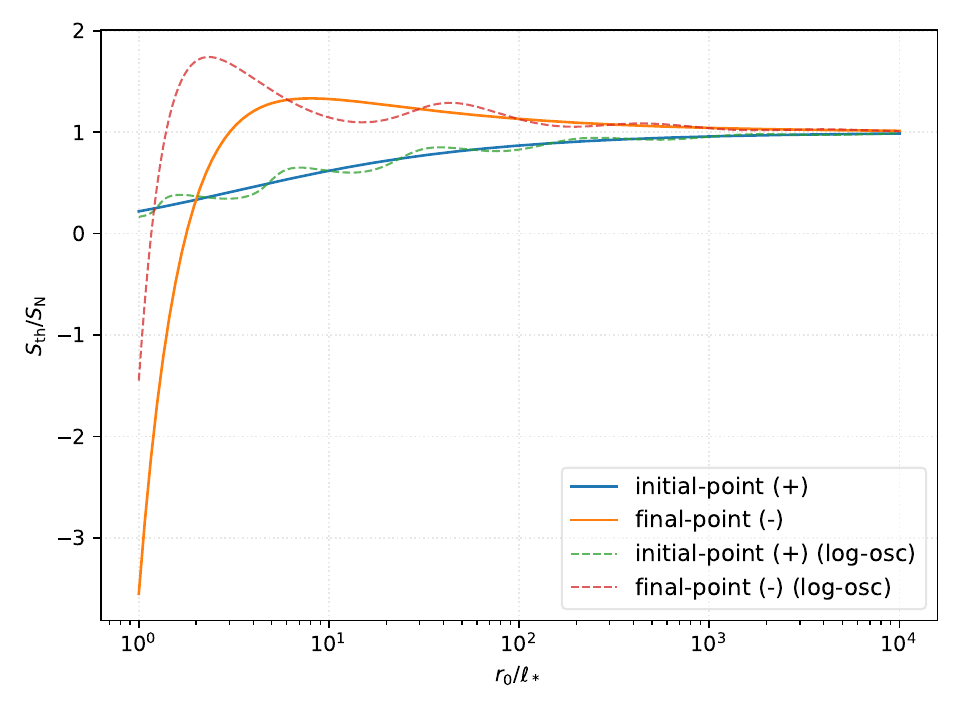}
    \caption{Integrable thermodynamic entropy relative to the Noether/area entropy, $S_{\rm th}/S_{\rm N}$, versus $r_0/\ell_*$ for the binomial profile (thick curves) in the two presentations. Thin dashed curves show a representative log-oscillating profile with small amplitudes. Here, $\alpha = 0.50$, and $\ell_* = 1.0$.}
    \label{fig2}
\end{figure}
Figure \ref{fig2} illustrates that the integrable entropy required by the fractional-frame temperature generally differs from the Noether entropy even though both are evaluated for the same geometric parameter $r_0$. For the binomial profile, $S_{\rm th}/S_{\rm N}$ deviates from unity at small $r_0/\ell_*$ and approaches $1$ in the infrared, consistently with the recovery of $q(r)\to r$. When logarithmic oscillations are included, the ratio acquires a bounded log-periodic modulation inherited from the profile through the integral definition of $S_{\rm th}$, while remaining insensitive to oscillations in the canonical quantity $S_{\rm N}=\pi r_0^2/G$. The plotted branches are restricted to configurations with $q'(r_{\rm h})>0$ so that the exterior branch is monotonic and the fractional-frame temperature is positive.

The corresponding multi-fractional potential conjugate to $\ell_*$ follows from Eq. \eqref{5.11}. Since
\begin{equation}
\partial_{\ell_*}q(r)=\pm \frac{1-\alpha}{\alpha}\,\ell_*^{-\alpha}\,r^\alpha,
\label{5.14}
\end{equation}
we obtain
\begin{equation}
\Psi_{\ell_*}
=\pm \frac{1-\alpha}{2G\,\alpha}\,\ell_*^{-\alpha}\,r_{\rm h}^{\alpha}
\left[
1-\frac{q'(r_{\rm h})}{q(r_{\rm h})}\,\frac{r_{\rm h}}{\alpha+1}
\right],
\label{5.15}
\end{equation}
where $q(r_{\rm h})=r_0$ and $q'(r_{\rm h})=1\pm (r_{\rm h}/\ell_*)^{\alpha-1}$ for the binomial profile.
\begin{figure}
    \centering
    \includegraphics[width=0.6\linewidth]{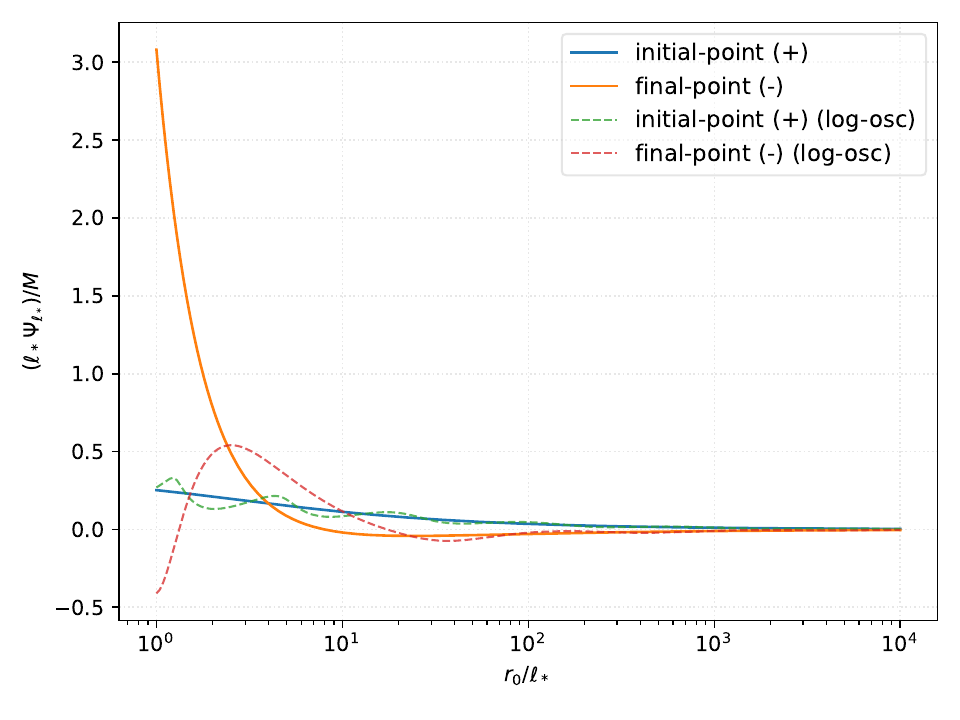}
    \caption{Dimensionless multifractional work potential $(\ell_*\,\Psi_{\ell_*})/M$ versus $r_0/\ell_*$ for the binomial profile (thick curves) in the two presentations. Thin dashed curves show a representative log-oscillating profile with small amplitudes. Here, $\alpha = 0.50$, and $\ell_* = 1.0$.}
    \label{fig3}
\end{figure}
Figure \ref{fig3} shows that varying the external scale $\ell_*$ induces a nontrivial generalized force $\Psi_{\ell_*}$ even though the canonical mass remains $M=r_0/(2G)$. For the binomial profile, $(\ell_*\,\Psi_{\ell_*})/M$ is appreciable only when $r_0/\ell_*$ is not large, and it decays in the infrared where $q(r)\to r$ and the multifractional corrections become negligible. When logarithmic oscillations are included, the same quantity acquires a bounded log-periodic modulation inherited from the profile dependence of $T_{\rm h}$ and of the integrable entropy $S_{\rm th}$, while the plotted branches are restricted to configurations with $q'(r_{\rm h})>0$ so that the exterior map is monotonic and the fractional-frame temperature is positive.

Analogous expressions are obtained for $\Psi_{\alpha}$ and, when logarithmic oscillations are present, for the additional potentials conjugate to $(A,B,\omega,\ell_\infty)$ by applying Eq. \eqref{5.11} to the profile \eqref{2.10}. In all cases, the work terms vanish when the profile is held fixed, and Eq. \eqref{5.10} reduces to the integrable fixed-profile relation \eqref{5.5}.

\section{Presentation dependence and oscillatory measures} \label{sec6}
The multi-fractional profile $q(r)$ is not a dynamical field but a fixed input specifying the theory. As a consequence, physical observables expressed in the fractional coordinate $r$ inherit a dependence on the choice of \emph{presentation} of the profile, in particular on the overall sign choice in the binomial deformation and on the phase of possible logarithmic oscillations \cite{Calcagni:2017ymp}. In the present context this dependence is entirely controlled by the map $q(r)$ evaluated at and near the horizon, because the canonical charges $(M,S_{\rm N})$ depend only on the geometric Schwarzschild parameter $r_0$, while the physical temperature $T_{\rm h}$ and the integrable thermodynamic entropy $S_{\rm th}$ depend on $q'(r_{\rm h})$ and on the integral of $q$ along $[0,r_{\rm h}]$ (Secs. \ref{sec3}--\ref{sec5}).

We first isolate the effect of the binomial presentation. For the coarse-grained profile
\begin{equation}
q(r)=r\pm \frac{\ell_*^{\,1-\alpha}}{\alpha}\,r^\alpha,
\qquad 0<\alpha<1,
\label{6.1}
\end{equation}
the horizon condition $q(r_{\rm h})=r_0$ becomes Eq. \eqref{2.16}. In the infrared regime $r_0\gg \ell_*$ (equivalently $r_{\rm h}\gg \ell_*$), it is useful to solve Eq. \eqref{2.16} perturbatively for $r_{\rm h}$ at fixed $r_0$. Writing $c=\ell_*^{\,1-\alpha}/\alpha$, one finds to leading order
\begin{equation}
r_{\rm h}=r_0\mp c\,r_0^\alpha+\mathcal{O}\!\left(\ell_*^{\,2(1-\alpha)}r_0^{2\alpha-1}\right),
\qquad (r_0\gg \ell_*),
\label{6.2}
\end{equation}
so that the two presentations correspond to two nearby fractional-frame horizon locations for the same geometric parameter $r_0$. The physical temperature in Eq. \eqref{3.5} becomes
\begin{equation}
T_{\rm h}
=\frac{1}{4\pi r_0}\left[1\pm\left(\frac{r_{\rm h}}{\ell_*}\right)^{\alpha-1}\right],
\label{6.3}
\end{equation}
and therefore, for large horizons, the presentation dependence is suppressed by the small factor $(r_{\rm h}/\ell_*)^{\alpha-1}$. The thermodynamic entropy defined by Eq. \eqref{5.13} exhibits the analogous behavior: for fixed $r_0$ the difference between the $+$ and $-$ presentations enters through the multifractional correction $\propto r_{\rm h}^{\alpha+1}$ and is likewise negligible in the infrared.

A useful way to interpret presentation dependence, advocated in Ref. \cite{Calcagni:2017ymp}, is to regard the presentation as an effective parametrization of an underlying microscopic (possibly stochastic) structure. This perspective is consistent with the broader association between dimensional flow and an effectively stochastic spacetime fuzziness discussed in Ref. \cite{Calcagni:2017jtf}. In that view one does not assign an operational preference to either sign in Eq. \eqref{6.1}, but instead treats the induced spread in fractional-frame observables as an intrinsic uncertainty band \cite{Calcagni:2017ymp}. For any quantity $\mathcal O$ that depends on the profile choice, one may then define the presentation average and half-spread as
\begin{align}
&\langle \mathcal O\rangle_{\pm}=\frac{\mathcal O_{+}+\mathcal O_{-}}{2},
\nn \\
&\Delta_{\pm}\mathcal O=\frac{|\mathcal O_{+}-\mathcal O_{-}|}{2},
\label{6.4}
\end{align}
where the subscripts indicate evaluation with the $+$ or $-$ presentation at fixed $r_0$. Since the canonical charges are presentation-independent, one has $\Delta_\pm M=\Delta_\pm S_{\rm N}=0$, while $\Delta_\pm T_{\rm h}\neq 0$ and $\Delta_\pm S_{\rm th}\neq 0$ in general. In the infrared, Eq. \eqref{6.3} implies $\langle T_{\rm h}\rangle_\pm\simeq (4\pi r_0)^{-1}$ and $\Delta_\pm T_{\rm h}\simeq (4\pi r_0)^{-1}(r_{\rm h}/\ell_*)^{\alpha-1}$, consistently with the recovery of the Schwarzschild temperature as an average value.

We now turn to logarithmic oscillations. The oscillatory profile in Eq. \eqref{2.10} can be written as
\begin{align}
&q(r)=r\pm \frac{\ell_*^{\,1-\alpha}}{\alpha}\,r^\alpha\,F_\omega(r),
\nn \\
&F_\omega(r)=1+A\cos\theta(r)+B\sin\theta(r),
\nn \\
&\theta(r)=\omega\ln\!\left(\frac{r}{\ell_\infty}\right),
\label{6.5}
\end{align}
with bounded amplitudes $A,B$ and dimensionless frequency $\omega$ \cite{Calcagni:2017ymp}. The physical temperature is still given by $T_{\rm h}=q'(r_{\rm h})/(4\pi r_0)$, but now depends on the oscillatory modulation through $q'(r_{\rm h})$. Using Eq. \eqref{3.7}, one sees that $T_{\rm h}$ contains both the value $F_\omega(r_{\rm h})$ and the phase-sensitive contribution from $F_\omega'(r_{\rm h})$. Consequently, $T_{\rm h}$ exhibits log-periodic oscillations as a function of $r_{\rm h}$, and hence as a function of $r_0$ through the implicit horizon equation. This behavior reflects discrete scale invariance \cite{Sornette:1997pb}, the log-periodic structure of fractal functions \cite{GluzmanSornette:2002logperiodic}, and the complex-dimension interpretation of the multi-fractional measure \cite{Calcagni:2017via}. The exterior-branch requirements stated in Sec. \ref{sec2}, in particular $q'(r)>0$ for $r\ge r_{\rm h}$, constrain the allowed amplitudes so that the oscillations do not generate turning points and additional horizons in the exterior region \cite{Calcagni:2017ymp}.

The thermodynamic entropy $S_{\rm th}$ defined by Eq. \eqref{5.7} can be expressed in closed form for the oscillatory profile because it depends only on $q(r)$, not on $q'(r)$. Substituting Eq. \eqref{6.5} into Eq. \eqref{5.7} gives
\begin{equation}
S_{\rm th}
=\frac{2\pi}{G}\int_{0}^{r_{\rm h}}\left[\bar r\pm \frac{\ell_*^{\,1-\alpha}}{\alpha}\,\bar r^\alpha\Big(1+A\cos\theta(\bar r)+B\sin\theta(\bar r)\Big)\right]d\bar r.
\label{6.6}
\end{equation}
The oscillatory integrals are elementary. Defining $\theta_{\rm h}=\theta(r_{\rm h})=\omega\ln(r_{\rm h}/\ell_\infty)$ and $K=(\alpha+1)^2+\omega^2$, one finds
\begin{align}
&\int_{0}^{r_{\rm h}}\bar r^\alpha \cos\theta(\bar r)\,d\bar r
=\frac{r_{\rm h}^{\alpha+1}}{K}\Big[(\alpha+1)\cos\theta_{\rm h}+\omega\sin\theta_{\rm h}\Big],
\nn \\
&\int_{0}^{r_{\rm h}}\bar r^\alpha \sin\theta(\bar r)\,d\bar r
=\frac{r_{\rm h}^{\alpha+1}}{K}\Big[(\alpha+1)\sin\theta_{\rm h}-\omega\cos\theta_{\rm h}\Big],
\label{6.7}
\end{align}
so that Eq. \eqref{6.6} yields a compact expression consisting of the binomial term in Eq. \eqref{5.13} plus an explicitly log-periodic correction proportional to $A$ and $B$. In particular, the oscillatory part is suppressed in the infrared by the same factor $r_{\rm h}^{\alpha+1}$ relative to the leading $\propto r_{\rm h}^2$ term, and it averages to zero if one averages uniformly over the phase $\theta_{\rm h}$.

The extended first law derived in Sec. \ref{sec5} remains valid when oscillations are present, provided the state space is further enlarged to include $(A,B,\omega,\ell_\infty)$ among the external parameters $\lambda^I$. The conjugate potentials are still given by the general formula \eqref{5.11}, with the required parameter derivatives read off directly from Eq. \eqref{6.5}, for example
\begin{align}
&\partial_A q(r)=\pm \frac{\ell_*^{\,1-\alpha}}{\alpha}\,r^\alpha\cos\theta(r),
\nn \\
&\partial_B q(r)=\pm \frac{\ell_*^{\,1-\alpha}}{\alpha}\,r^\alpha\sin\theta(r),
\label{6.8}
\end{align}
and similarly for $\partial_\omega q$ and $\partial_{\ell_\infty}q$ through the derivatives of $\theta(r)$. The resulting $\Psi_I$ are phase-dependent and inherit the same log-periodic structure as $T_{\rm h}$, while the canonical mass $M=r_0/(2G)$ and Noether entropy $S_{\rm N}=\pi r_0^2/G$ remain unaffected by oscillations at fixed $r_0$. This separation between profile-sensitive thermal quantities $(T_{\rm h},S_{\rm th},\Psi_I)$ and profile-insensitive canonical charges $(M,S_{\rm N})$ is the operational content of presentation dependence in the thermodynamic language \cite{Calcagni:2017ymp}.

This distinction also clarifies how the present construction should be interpreted in relation to possible no-hair or uniqueness statements. The parameters $\lambda^I$ are not integration constants of a new black-hole solution and are not dynamical fields carrying independent asymptotic charges. They specify the external multi-fractional profile $q(r,\lambda)$. Thus, at fixed profile, the static vacuum solution considered here remains the Schwarzschild family in the geometric areal coordinate $q$, and its canonical charges are still labeled only by $r_0$. Allowing variations of $\lambda^I$ in the thermodynamic first law should therefore be viewed as comparing neighboring profiles, presentations, or theory sectors, not as introducing additional black-hole hair in the usual sense. Nevertheless, the profile dependence of fractional-frame observables suggests that any future no-hair or uniqueness theorem in multi-fractional gravity must state explicitly whether the profile $q(r,\lambda)$ is held fixed as part of the theory data or is included in a broader classification of admissible backgrounds.

\section{Conclusion} \label{sec7}
We considered the static, spherically symmetric vacuum sector of the multi-fractional theory with $q$-derivatives in the spherical-coordinate approximation, where the geometry is Schwarzschild in the geometric areal radius $q$ and is pulled back to the fractional radial coordinate $r$ through a prescribed multi-scale map $q=q(r)$ \cite{Calcagni:2017ymp}. The central point is that canonical charges are defined covariantly in the geometric frame. As a consequence, the conserved mass is fixed solely by the Schwarzschild integration constant $r_0$ through Eq. \eqref{4.9}, and the Noether (Iyer--Wald) entropy is the standard area law in terms of the geometric horizon radius $q(r_{\rm h})=r_0$ through Eq. \eqref{4.13}. These canonical quantities are insensitive to variations of the non-dynamical profile parameters entering $q(r)$ at fixed $r_0$.

In contrast, if one adopts the fractional-frame definition of the Hawking temperature used in the multi-fractional black-hole literature \cite{Calcagni:2017ymp}, the temperature depends on the local weight factor at the horizon, $T_{\rm h}=q'(r_{\rm h})/(4\pi r_0)$ in Eq. \eqref{3.5}. Allowing variations of the profile parameters then makes $T_{\rm h}$ vary even when the canonical mass does not, which obstructs a naive identification of an entropy via $dS=dM/T_{\rm h}$. We resolved this obstruction by enlarging the thermodynamic state space to include the profile parameters $\lambda^I$ and by constructing an integrable entropy state function
\begin{equation}
S_{\rm th}=\frac{2\pi}{G}\int_{0}^{r_{\rm h}} q(\bar r,\lambda)\,d\bar r,
\label{7.1}
\end{equation}
which reduces to the Noether entropy in the general-relativistic limit. On this extended space the first law takes the integrable form
\begin{equation}
dM=T_{\rm h}\,dS_{\rm th}+\Psi_I\,d\lambda^I,
\label{7.2}
\end{equation}
with uniquely determined multi-fractional potentials $\Psi_I$ given by Eq. \eqref{5.11}. In this sense, insisting on the physical temperature prescription in the fractional coordinate forces additional thermodynamic work terms when theory parameters are varied.

The same logic clarifies the role of presentation dependence. For the binomial profile, the $+$ and $-$ choices correspond to distinct fixed profiles, hence to distinct thermal quantities $(T_{\rm h},S_{\rm th},\Psi_I)$ at the same geometric parameter $r_0$, while leaving $(M,S_{\rm N})$ unchanged. For log-oscillating profiles, the thermodynamic quantities inherit a phase dependence controlled by $\omega$ and $(A,B)$, again without affecting the canonical charges at fixed $r_0$. One may either treat these differences as genuinely distinct theories (different choices of $q$) or, following the stochastic interpretation discussed in Ref. \cite{Calcagni:2017ymp}, regard them as an uncertainty band in fractional-frame thermodynamic observables.

We collect here the consistency conditions that delimit the regime where the preceding thermodynamic formulation is meaningful. First, the exterior region must be well defined. In practice, we restrict to an exterior branch satisfying $q(r)>0$ for all $r\ge r_{\rm h}$, so that the areal radius is real and the angular sector is non-degenerate. Second, the map must be monotonic in the exterior, $q'(r)>0$ for all $r\ge r_{\rm h}$, so that the geometric radius is a valid coordinate and the temperature \eqref{3.5} is positive. Third, the horizon equation $q(r_{\rm h})=r_0$ must admit a unique solution on the chosen exterior branch, ensuring that horizon observables are single-valued functions of the parameters. In the binomial final-point presentation, the additional finite-radius zero of $q(r)$ in Eq. \eqref{2.17} lies at $r\sim\ell_*$ and signals that one must choose the outer monotonic branch when discussing asymptotically flat exteriors. With logarithmic oscillations, large amplitudes can generate additional zeros or turning points at ultra-short distances \cite{Calcagni:2017ymp}; the above monotonicity and uniqueness requirements impose corresponding bounds on $(A,B)$ (for fixed $\omega$ and $\ell_\infty$) in any application where a single-horizon exterior geometry is assumed.

A natural direction for further investigation is to apply the same extended-state-space logic to more general solution families in the $q$-derivative theory, such as charged or (within an appropriate approximation) rotating configurations, and to analyze whether the separation between profile-insensitive canonical charges and profile-sensitive thermal response persists in those settings, always under the same exterior-branch consistency conditions. This question is closely related to possible no-hair and uniqueness statements in multi-fractional gravity. The present results suggest that, at fixed profile $q(r,\lambda)$, the parameters $\lambda^I$ should be regarded as external theory data rather than as additional black-hole hair: the canonical solution in the static vacuum sector is still labeled by the Schwarzschild parameter $r_0$. If one enlarges the classification problem to include different admissible profiles, then different choices of $\lambda^I$ may lead to distinct fractional-frame observables and thermodynamic response functions even when the geometric-frame charges coincide. Establishing whether this profile sensitivity can mimic, modify, or evade conventional uniqueness expectations in charged, rotating, or less symmetric sectors remains an important problem for future work.

\acknowledgements
R. P. would like to acknowledge networking support of the COST Action CA21106 - COSMIC WISPers in the Dark Universe: Theory, astrophysics and experiments (CosmicWISPers), the COST Action CA22113 - Fundamental challenges in theoretical physics (THEORY-CHALLENGES), the COST Action CA21136 - Addressing observational tensions in cosmology with systematics and fundamental physics (CosmoVerse), the COST Action CA23130 - Bridging high and low energies in search of quantum gravity (BridgeQG), the COST Action CA23115 - Relativistic Quantum Information (RQI) funded by COST (European Cooperation in Science and Technology), and SCOAP3 (Switzerland) for their support.

\bibliography{ref}

\end{document}